# The application of a perceptron model to classify an individual's response to a proposed loading dose regimen of Warfarin.

Cen Wan[1], I.V.Biktasheva[1], S.Lane[2]

[1] Department of Computer Science, University of Liverpool, UK

[2] Department of Biostatistics, University of Liverpool, UK

What is already known about the subject

Warfarin is a drug with high inter-individual variability and although numerous studies have been undertaken to determine an individual's response to the dose given in the maintenance phase, little has been done to predict an individual's potential response to the initial loading regimen.

What this paper adds

The aim of the loading regimen is to bring the International Normalisation ratio into the required therapeutic range. This study proposes a method for pre-classifying an individual's response to the proposed loading regimen. The method takes into account other variables that may influence the response and with a high degree of accuracy classes the potential response as being either below therapeutic range, within therapeutic range or above therapeutic range

**Abstract**

*Background and purpose*

The dose regimen of Warfarin is separated into two phases. Firstly a loading dose is given, which is designed to bring the International Normalisation Ratio (INR) to within therapeutic range. Then a stable maintenance dose is given to maintain the INR within therapeutic range. In the United Kingdom (UK) the loading dose is usually given as three individual daily doses, the standard loading dose being 10mg on days one and two and 5mgs on day three, which can be varied at the discretion of the clinician. However, due to the large inter-individual variation in the response to Warfarin therapy, it is difficult to identify which patients will reach the narrow therapeutic window for target INR, and which will be above or below the therapeutic window. The aim of this research was to develop a methodology using a neural networks classification algorithm and data mining techniques to predict for a given loading dose and patient characteristics if the patient is more likely to achieve target INR or more likely to be above or below therapeutic range.

*Experimental Approach*

Multilayer perceptron (MLP) and 10-fold stratified cross validation algorithms were used to determine an artificial neural network to classify patients' response to their initial Warfarin loading dose.

*Key Results*

The resulting neural network model correctly classifies an individual's response to their Warfarin loading dose over 80% of the time. As well as taking into account the initial loading dose, the final model also includes demographic, genetic and a number of other potential confounding factors.

*Conclusion and implications*

With this model clinicians can predetermine whether a given loading regimen, along with specific patient characteristics will achieve a therapeutic response for a particular patient. Thus tailoring the loading dose regimen to meet the individual needs of the patient and reducing the risk of adverse drug reactions associated with Warfarin.

*Keywords*

Warfarin, Narrow therapeutic window, loading dose, neural networks

**Introduction**

Warfarin is an anticoagulant drug, which is the front line treatment for individuals at risk of thrombosis. Although Warfarin is effective in many cases, its effectiveness is subject to high inter-patient variability, with some patients requiring small doses, as low as 1mg per day and others relatively high doses, up to 20mg per day. Another drawback to the use of Warfarin is the high risk of adverse events, including haemorrhage. A study by Pirmohamed et al, 2004 showed that Warfarin is responsible for approximately 10% of all hospital admissions for adverse drug reactions. Hence the requirement to provide patients with an individualised dose that maximises the efficacy of the drug whilst at the same time minimising the risk of potential adverse reactions.

The International Normalisation Ratio (INR), a measure of the clotting tendency of blood is used to monitor a patient's response to the drug. The usual therapeutic range for INR is between 2 and 3; however, individuals with prosthetic heart values can have higher values between 3 and 4. The dosing regimen of Warfarin is divided into two phases. Initially, a 3 day loading phase is given to bring a patient's INR, measured on day 7, into the therapeutic range and then a maintenance dose is given to maintain the patients INR within therapeutic range. A number of algorithms have been proposed to estimate a patient's maintenance dose (e.g. Gage et al, 2008, Schelleman, 2008, Zhu, 2007, Sconce, 2005). The International Warfarin Pharmacogenetics Consortium used an ordinary least-squares linear regression modelling method to develop a pharmacogenetic algorithm that predicts the square root of the dose and incorporates both genetic and clinical data (The International Warfarin Pharmacogenetics Consortium, 2009). They found that greatest benefits from inclusion of the genetic profiles into their model were observed for the two extreme groups of patients: those in the 46.2% of the population that required 21 mg or less of warfarin per week and those that required 49 mg or more per week for therapeutic anticoagulation. In (McDonald et al, 2008), using data on maintenance dose response for 17396 patients, the authors performed a comparison study of three types of model: polynomial, auto-regression/moving average with exogenous variable (ARMAX) (Hannan, 1976), and neural network (NN) in predicting INR values. Using historical data on patients' INR measuments, though no genetic data, McDonald et al showed that continuously-updating neural network models predicted future INR measurements best of the three types of models. However, there have been few studies investigating the response of patient's to specific loading dose regimens. In the UK loading doses are normally three daily doses of Warfarin, the most common being 10mgs on days one and two and 5mgs on day three. However, this can be varied at the discretion of the clinician. In particular, lower doses, e.g. 5mgs on each of the three days, can be prescribed if the patient is thought to be frail or has an increased risk of bleeding. Previously unpublished work (Lane and Biktasheva, 2008) showed that patients given a conservative loading regimen were unlikely to achieve an INR within therapeutic range irrespective of their individual characteristics. As it is well known that a patient response depends on both patient's characteristics and the loading dosage, and some combinations of the two result in counterintuitive responses, it seems to be the main difficulty preventing the lineage of the Warfarin loading dosage decision based on pure patients characteristics. At present, there is no optimal approach for determining the most suitable loading regimen for a specific patient.

A patient's response to Warfarin can be confounded by a number of factors. Demographic factors such as age, race or gender (Takahashi et al, 2006, Kamali *et al*, 2004) can have a significant impact on how the body responds to the prescribed dose, other factors such as co-medications, e.g. Amiodarone, and diet; particularly vitamin K intake can also influence response (Wells et al, 1994, Ansell et al, 2008). More recently genetic variations have also been identified, particularly in the genotypes CYP2C9 and VKORC1, which have a major impact on how the body responds to Warfarin therapy (Zhu et al, 2007).

Regression models make use of statistical analysis to assess significance of each covariate proposed for the model in terms of how they affect the outcome of interest so non-significant (unless deemed clinically important) covariates can be excluded from the model. Then, given a set of measurements the regression models provide a prognostic formula for predicting a specific outcome value or probability of a specific response when the outcome is binary or categorical. This makes regression models easy to understand and interpret by clinicians. However, linear regression has a significant drawback: it does not work well for nonlinear problems, so to improve performance it requires fine tuning of nonlinear kernels, adaptive regression splines, etc. Published regression algorithms often explain less than 50% of the variability in the outcome variable, particularly when applied to datasets other than the one used to derive the algorithm (International Warfarin Pharmacogenetics Consortium, Klein et al, 2009). A neural network approach implicitly detects all complex nonlinear relationships between independent and dependent variables, and is remarkably resilient to the noise always present in large data sets. Also, as a pattern recognition tool, neural network can provide classification of the response when more than two classes are present. It has been demonstrated that neural networks can be applied successfully for Warfarin maintenance dose prediction (Solomon et al, 2004) and to predict the daily INR level of individual patients (Narayaman et al, 1993). The results were promising, but further investigation was needed.

The objective of the study reported in this manuscript was to apply data-mining techniques and neural network models to obtain high accuracy classification of patients' response to their Warfarin loading dose regimen, taking into account demographic, genetic and other measurable confounding factors. The aim was to identify which patients would achieve the therapeutic INR range during the initial loading stage, which would be below the required threshold and which would be above the therapeutic window, that is to predict the three possible types of individual response: "in therapeutic range", "under load", and "over load". We were also interested in determining if additional INR measurements on day 4, 5, and 6, which currently are not in clinical practice but were available for this study, can improve predictability of an individual patient's response to Warfarin loading regimen.

Information on 500 patients taking Warfarin was available for the study. Data was collected on the patients' demographic and genetic variables, as well as on whether or not they were taking the antiarrhythmic drug Amiodarone, their loading dose regimen, baseline INR, and INR measurements on days 4, 5, 6 and 7. The stratified cross validation data mining technique was used to ensure that the patterns of response to Warfarin loading dose regimen are of general application.

**Methods**

Data from patients who had started Warfarin treatment, irrespective of indication, between November 2004 and March 2006 were included in the analysis after having obtained written informed consent. A total of 500 patients were recruited from the Royal Liverpool and Broadgreen University Hospital NHS Trust and University Hospital Aintree, Liverpool, UK. All patients were informed of the nature of the study prior to being enrolled and the only exclusion criteria was the refusal to give written informed consent. The study was approved by Birmingham South Research Ethics Committee. The mean age of the cohort was 68.36 years, 275 (54.9%) were male and 45 (9.0%) were also taking the drug Amiodarone. Two genes VKORC1 and CYP2C9 were genotyped. For the VKORC1 haplotype, 65 (13%) were GG, 189 (37.7%) were AA and 208 (41.5%) were 41.5% the remainder were unknown (7.8%). For the CYP2C9 genotype, 347 (69.3%) were the wild type *1/*1, 96 (19.2%) were *1/*2, 42 (8.4%) were *1/*3, 4 (0.8%) were *2/*2, 8 (1.6%) were *2/*3 and 3 (0.6%) were *3/*3. One-hundred and thirty-three (45%) were prescribed the standard loading regimen of 10mg, 10mg, 5mg over the 3 days, 19 (6.5%) had a loading dose regimen of 10mg, 10mg, 10mg, similarly 19 (6.5%) had a loading dose regimen of 7mg,7mg,7mg, 15 (5.1%) had a loading dose of 5mg,5mg,5mg, and the remaining 36.9% had loading dose regimens ranging from 0mg, 7mg, 7mg to 10mg, 10mg, 12mg.

The complete list of the covariates is as follows: age (yrs), gender (male=1, female=0), body-weight (kg), height (metres), BSA, BMI, Amiodarone (yes=1, no=0), VKORC1 (GG=0, AA=1, AG=2), CYP2C9 (*1/*1=0, *1/*2=1, *1/*3=2, 2*/*2=3, 2*/*3=4, *3/*3=5), loading dose day1, loading dose day 2, loading dose day 3, INR day 4, INR day 5, INR day 6 and baseline INR. The total loading dose, equals sum of loading doses 1, 2 and 3 was later included in the model.

Prior to applying the Perceptron algorithm to the dataset, some pre-processing was required to deal with missing data. The data pre-processing/cleaning steps are illustrated in Figure. 1.

Firstly, individuals missing all body measurements (height, weight, BMI) and demographic data (sex, age) were excluded from the dataset. Then, individuals missing all loading dose information on days 1-3 were excluded from the dataset. Due to clinics being closed on weekends etc., very few individuals had a full complement of the INR measurements on days 4, 5, 6, with a small minority having missing INR measurements on day 7. Consequently, missing INR values on day 7 were replaced by the value nearest to day 7 that is by the INR value on day 8. If missing both days 7 and 8, then INR values on day 7 were replaced by the INR value on day 6. The day 7 INR values were then categorised as follows; response below target (categorised 1), in-range (categorised 2), and above target (categorised 3). After that, the individuals with all missing interim measurements of INR day 4, 5, and 6 were excluded. Finally the remaining missing values were replaced with 0.

The final dataset used in classification experiments contained 294 individual observations, on 16 explanatory variables and one outcome variable, the categorised response to the given loading dose. Later, the total loading was included the model as an extra explanatory variable.

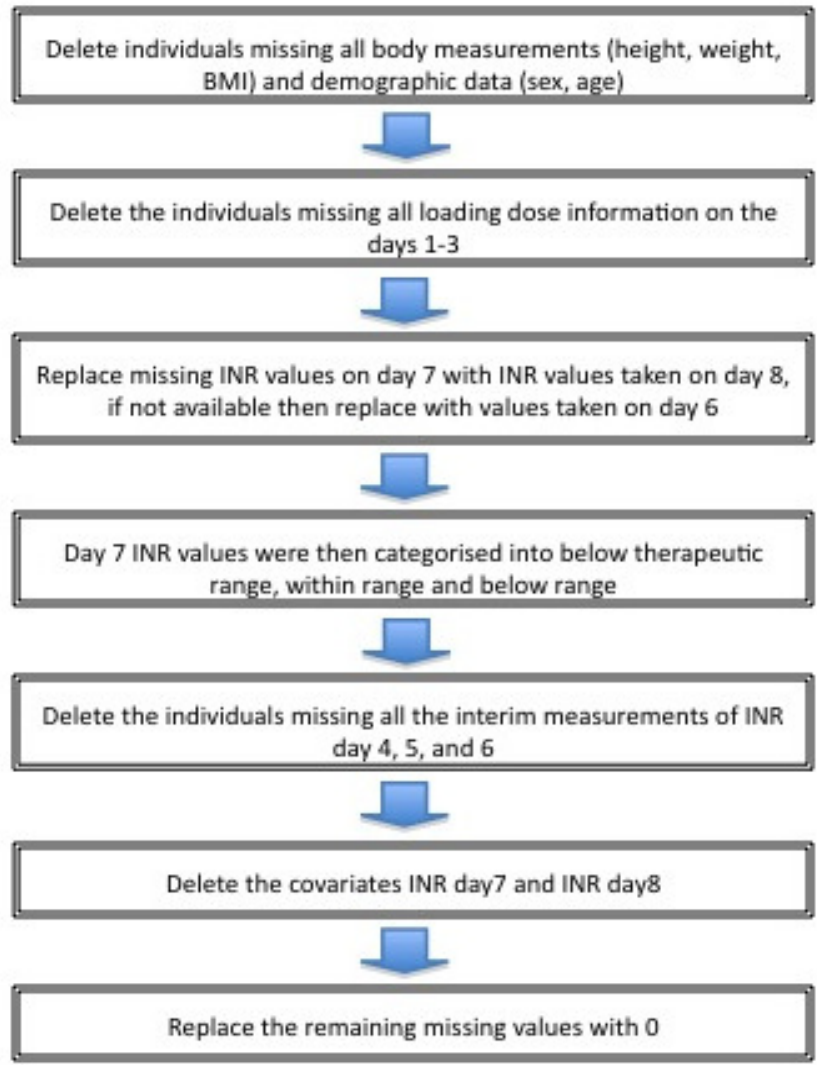

**Figure 1:** Data pre-processing flow.

A Matlab implementation of the Multilayer Perceptron, MLP (Haykin, 1999) was used to construct the classification algorithm. A very useful and concise explanation of the Perceptron algorithm can be found in the Appendix of Solomon et al, 2004. The artificial neural network uses a training data set to learn the mapping of input instances into known "class" outputs. After the mapping is established, it can be used to predict classification of new data. The data set, described above, was used to train the MLP to map the input instances of patients' individual data into their known individual categorised INR day 7 taken as "class" output. After the iterative training, the most appropriate weights for the input data are obtained, and the algorithm is able to provide accurate classification predictions for new patients.

Although the MLP algorithm is well developed and wildly used for classification problems, its success is highly dependent on the quality/representativeness of the training data and a good choice of the MLP architecture such as number of hidden layers/neurons and choice of activation functions, which are in turn problem dependent, so in practice it has to be defined empirically by trial and error.

The models were validated on independent data using stratified 10-fold cross validation (Witten and Frank, 2005). As the standard loading regimen of Warfarin is varied at the discretion of an experienced clinician, there is a bias in the data set towards the majority of the patients who reach the therapeutic INR on day 7, however, the main challenge is to identify those patients who have the risk of under- or over-load. The random selection of records into the training and validation/test sets does not take into account the unequal distribution of response categories, thus failing to improve classification beyond 60% (D. Fillingham, 2011). That is why stratification of the independent testing set is necessary to preserve the same representativeness proportion of each category as in the training set. Following the data mining 10-fold stratified cross-validation technique, the original dataset is divided into 10 subsets, with the proportion of categories in each subset being the same as the proportion of categories in the whole sample. Then 9 of these subsets are used for training with the 10th subset used for validation. Then, each of the 10 subsets is used in turn as the test set with the rest of the data being used for training. Finally, after 10 runs the average accuracy is obtained. Thus, the 10-fold stratified cross validation ensures that the models are validated on independent data.

Both the Multilayer Perceptron and the 10-fold stratified cross validation were implemented using MATLAB version 7.8.0.

**Results**

**Optimal Structure of Multilayer Perceptron**

The first part of the study was to identify the best structure for the Multilayer Perceptron model for the given data set. The optimal MLP structure was defined by conducting 10-fold stratified classification experiments on the pre-processed dataset described above. The outcome, "class" variable that is the categorised value of INR day7 was represented by a single node in the Perceptron output layer, coded as class "1" for "under load", class "2" for "in therapeutic range", and class "3" for "over load". The number of hidden layers/nodes, learning rate, number of training iterations, as well as combination of activation functions for the hidden and output nodes were varied to define the optimal Perceptron structure to produce the best accuracy classification.

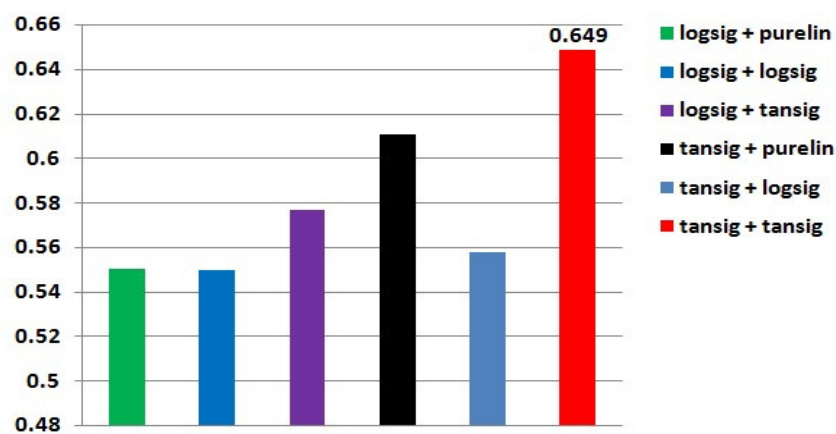

**Figure 2**: Average classification accuracy obtained by different activation functions.

First, the best combination of activation functions for hidden and output layer nodes was defined for a Perceptron with a single hidden layer consisting of 5 hidden nodes, a fixed learning rate of 0.15 was set and with a run of 100 iterations. The average accuracy obtained from different combinations of activation functions is shown in Figure 2, where "purelin" is the pure linear activation function, "logsig" is logarithmic sigmoid activation function and "tansig" is tangent sigmoid activation function. The default Matlab setting is the logarithmic sigmoid activation function for hidden neurons with the pure linear activation function for the output neurons.

It was found that the tangent sigmoid function is the best activation function for both the hidden and output layers, as it gives an average accuracy rate of 0.649 for the data set. Further experiments to determine the best learning rate and corresponding iteration times identified 0.04 as the best learning rate and 3000 to be the optimal number of iterations. With this learning rate and number of iterations, the average accuracy was increased to 0.835. Finally, further experiments were carried out to investigate whether increasing the number of hidden layers/nodes would improve the accuracy rate; however it proved not to be the case. The final perceptron model contained a single hidden layer with 5 nodes, with the tangent sigmoid activation function use in all hidden and output nodes, the learning rate was 0.04, and was run for 3000 iterations.

**Data Quality.**

To appreciate the importance of data quality, classification using the pre-processed dataset and the original dataset with missing data was conducted. In the original data set, the instances with missing INR values on day 7 were labelled as an additional category 4, all other missing values were replaced with 0. Missing data represents a particular type of noise in a data set and different methodological techniques can be used to deal with it. For continuous numerical covariates it may be sensible to replace missing data with an average or average for the "class". However, it is not possible to replace categorical missing data with an average; for example, sex, or genotype, even average age may not make sense as the noise will still be there. From this point of view simple replacement of a missing data with 0, is just as plausible as labelling it as "noise" in the pattern, and quantify the effect of the noise on classification than trying to smooth it out. The classification outcomes using the pre-processed dataset and the original dataset with missing data are shown in Figure 3. The confusion matrices are used to show the correctly identified number of instances as "true" class on the main diagonal, with misclassified "false" class instances above and below the main diagonal. Note as 10-fold cross validation was used for all classification experiments, the average number of instances for each "true" and "false" classification is shown in the confusion matrices rather than counts of individuals, hence the reporting of decimal values. For example, in Figure 3a (line 1), there is an average of 55.2 patients in category 1 correctly identified using the 10-fold cross validation, average of 13.8 patients of category 2 patients wrongly identified as being in category 1, and average of 1.2 category 3 patients wrongly identified as being in category 1. The average percentage of correctly identified patients (e.g. 78.6% for class 1) is given in column 4. Cell 2 in column 1 shows average number of patients in category 1 wrongly classified as category 2. Cell 3 in column 1 shows average number of patients in category 1 wrongly classified as category 3. The bottom cell 4 in column 1 shows average percentage (78.9%) of correctly identified/recalled patients in category 1. Also, in Figure 3b, as all instances of class 4 that is "missing INR day 7" had to be identified as either class 1, 2, or 3, there is empty "true" cell for the class 4.

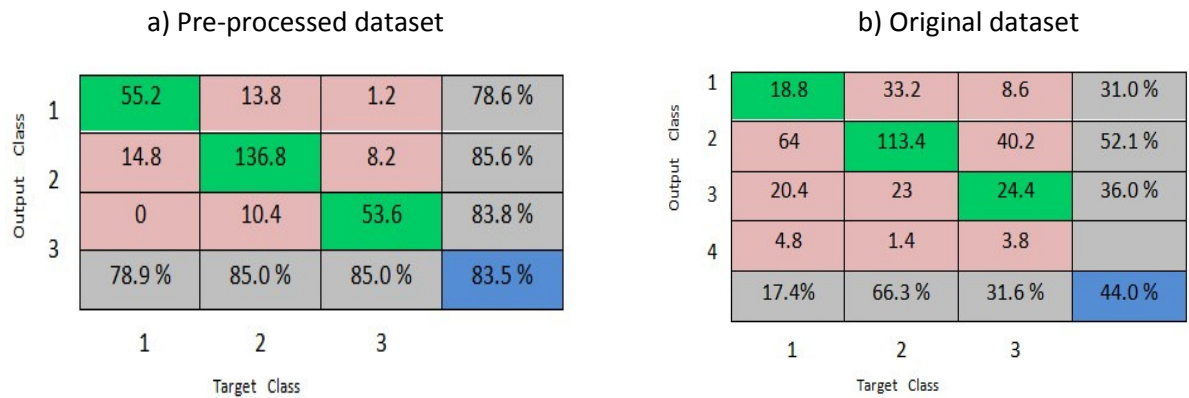

**Figure 3**: Confusion matrix for classification using a) pre-processed and b) original dataset.

According to the confusion matrices in Figure 3, the classification performance of the pre-processed dataset is much better than the original dataset which included the missing data. Especially for class 1, the under-loads, the accuracy rate for the original dataset is only 17.4% whereas the pre-processed dataset has an accuracy rate of 78.9%. Although the Perceptron algorithm has the ability for dealing with missing values, up to some critical amount, missing values have a negative influence on the accuracy rate. The results demonstrate that the percentage of missing values in the unprocessed original data set was high, resulting in poor classification performance, particularly on the classes which contain fewer observations.

**Optimal set of covariates.**

For the reasons discussed above, in the original data set, and even after pre-processing, there were many missing interim INR measurements on days 4, 5, and 6, with particularly scarce data on INR day 5 and 6. On the other hand, as our unpublished work (Lane and Biktasheva, 2008) showed patients given a too conservative loading regimen were unlikely to achieve an on INR day 7 within the therapeutic range, it was important to investigate the classification effect of the total loading, consequently this measurement was included in the analysis as an additional covariate. Therefore, five possible combinations of covariates in the data set, shown in Table 1, with presence/absence of total loading and the interim INR measurements on days 4, 5, and 6 were tested for classification outcome.

| Data Set № | INR day 4 | INR day 5 and 6 | Total Loading | Comments |
|---|---|---|---|---|
| 1 | + | + | - | The original pre-processed data set |
| 2 | + | - | - | |
| 3 | - | - | - | |
| 4 | + | - | + | |
| 5 | - | - | + | Optimal, best classification |

**Table 1:** Data sets used to identify optimal combination of covariates.

To determine what role the infrequently measured covariates, INR day 4, 5, and 6, might have on the classification results, first, INR day 5 and day 6 were deleted from the pre-processed data set (Data

Set 2 in the Table1), and the experiments were repeated on this reduced number of covariates. Then, the experiments were repeated with the pre-processed data set with all three covariates INR day 4, 5, and 6 deleted (Data Set 3 in the Table1). The classification results are shown in Figure 4.

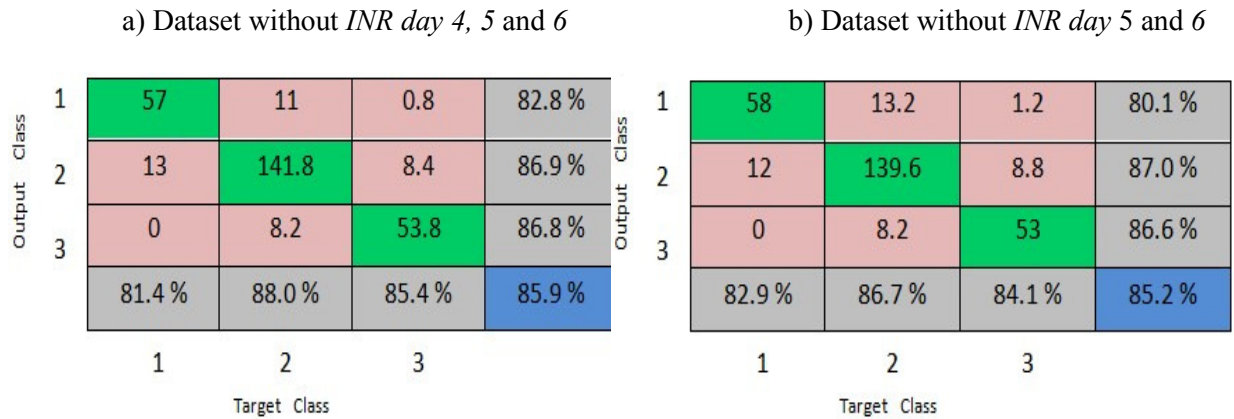

a) Dataset without *INR day 4, 5 and 6*     b) Dataset without *INR day 5 and 6*

**Figure 4:** Confusion matrices for classification with the a) data set 3, and b) data set 2, as in Table 1.

It can be seen by comparison of Figures 3a and 4b, that elimination of the interim measurements of INR day 5 and 6, improves classification of classes 1 and 2, the "under-load" and "in therapeutic range" categories, but reduces classification accuracy for class 3 "over-load". Further elimination of the covariate INR day 4, improves classification of classes 2 and 3, the "in therapeutic range" and "over-load" group, but again reduces classification accuracy for class 3 "over-load". Consequently, at this stage, it is difficult to say whether or not the classification improvements were due to the elimination of missing data, or the covariates INR day 5, 6, 7 which have some effect on classification accuracy for the classes having fewer observations. However, comparing Figures 3a and 4a, classification without INR day 4-6 is much better for all categories than the one obtained on the pre-processed data with the complete original set of covariates.

Finally, to check if the total loading dose improves classification, it was included the model as an extra covariate. The results are shown in Figure 5.

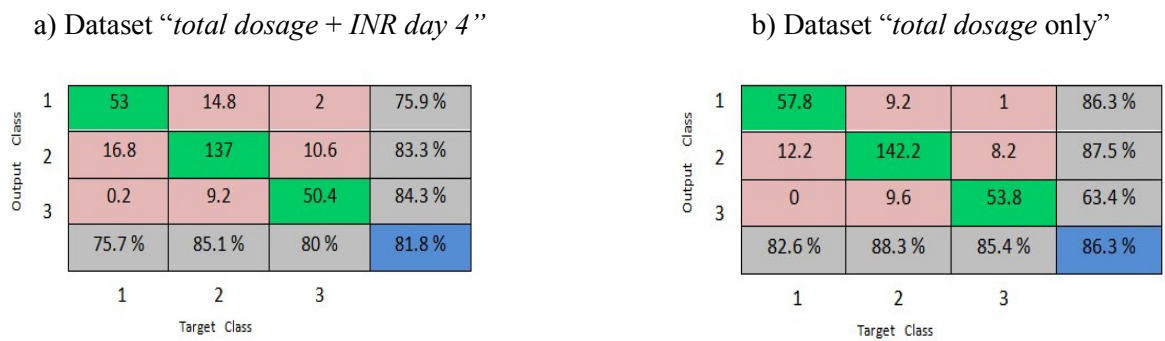

a) Dataset "*total dosage + INR day 4*"     b) Dataset "*total dosage* only"

**Figure 5:** Confusion matrices for classification experiments with datasets 4 and 5, as in Table 1.

From comparison of Figures 4a and 5b, it can be seen that adding the total loading dose to the data set 3 in Table 1, that is *without INR day 4-6*, further improves the classification rate for all three classes, so the average classification rate data set 5 in Table 1 is very high 86.3%, with 82.6% accuracy on the smaller "under-load" category 1, and 88.3% on the large "in therapeutic range" category 2.

From comparison of Figures 5a and 5b, it can be seen that again the addition of the covariant INR day4 has a significant negative effect on classification for all the three classes.

**Discussion**

Although one previous study, Solomon et al, 2004, has successfully applied neural network techniques to Warfarin maintenance dose prediction, and there have been a number of linear regression models proposed for estimating the maintenance dose, this is the first study that has investigated the individual's response to the initial loading dose phase. One major problem with estimating the response to the loading dose is that an individual INR level is subject to high daily variability, and in many cases individuals do not reach a steady state. However, due to the high risk of adverse drug reactions associated with Warfarin, it is important that both loading and maintenance dosing regimens are as individualised as possible. A number of demographic variables, genotypes, along with co-medication and diet have been identified as possible causes of the high inter-individual variability in dose requirements and response to the drug.

The Day 7 INR value was chosen as the output variable, because it was thought to provide the first indication of the individual's steady state INR value. Earlier measurements taken on days 4 or 5 are still subject to high fluctuations and the perceptron model demonstrated that when these measurements were included as explanatory variables they had a negative impact on the accuracy of the model. However, the model did demonstrate that by combining the individuals loading dose, in this case three daily doses of Warfarin, along with known demographic and genotype information it was possible to classify the individual's response at day 7 with a high degree of accuracy. This accuracy was improved further when the total three day loading dose was added to the model.

The initial accuracy of the model to correctly classify an individual's response to their prescribed loading dose was 78.9% for those who failed to achieve target INR, 85.0% for those who achieved the therapeutic range, and 85.0% for those who were above the therapeutic range. When total dosage was added to the model the accuracy in classification rates increased to 82.6%, 88.3% and 85.4% respectively.

Further development of the algorithm to predict the optimum loading dose for an individual to achieve the therapeutic range will be of a particular value to clinicians. This is to be the subject of a follow-up study.

**Conclusion**

The study demonstrates that the perceptron algorithm can be used to classify a patient's response to a specific loading dose regimen given demographic and genetic information. Consequently, at day 0 the clinician can test different loading dose strategies, (e.g. 10, 10, 5 or 8, 8, 7) with the known demographic and genetic information to predict the individual's response, in-terms of being below, within or above the required therapeutic range at day 7.

The stratified cross validation warrants the models testing on independent data and generalization of the results to the complete data set. The algorithm has good accuracy for classifying the patient's response and could be implemented into clinical practice, as an aid to ensuring that patients receive the best possible dose to achieve therapeutic INR and reduce the risk of adverse drug reactions.

The neural networks methods seem to take better account of this implicit nonlinear link between the patients characteristics, the Warfarin loading dosage and individual patients response.